\documentclass[aps,
prl,
reprint,
twocolumn,
]{revtex4-2}

\usepackage{amsmath}
\usepackage{graphicx}
\usepackage{bm}
\usepackage{hyperref}
\usepackage{xcolor}
\usepackage{nicefrac}

\usepackage{braket}
\usepackage{array}
\usepackage{dsfont}
\usepackage{pifont}
\usepackage[normalem]{ulem}
\usepackage{amssymb}

\usepackage{amsfonts}
\usepackage{mathtools}
\usepackage{graphicx}
\usepackage{dcolumn}
\usepackage{bm}
\usepackage{multirow}
\usepackage{hhline}
\usepackage{siunitx}

\usepackage{leftidx}
\usepackage{color}
\usepackage{mathtools}
\usepackage[mathscr]{eucal}
\usepackage[german,english]{babel}
\usepackage[capitalize]{cleveref}
\usepackage{orcidlink}

\DeclareUnicodeCharacter{2009}{ }

\newcommand{\vct}[1]{{\bf #1}}
\newcommand{\nf}[2]{{{#1}/{#2}}}

\begin{document}

\title{How dipolar interactions structure molecular droplets}

\author{William Freitas \orcidlink{0000-0002-8020-2117}}
\affiliation{%
Max Planck Institute for the Physics of Complex Systems, 
Noethnitzer Str. 38, 01187 Dresden, Germany}
\author{Panagiotis Giannakeas \orcidlink{0000-0002-1164-0201}}

\affiliation{%
Max Planck Institute for the Physics of Complex Systems, 
Noethnitzer Str. 38, 01187 Dresden, Germany}
\author{Jan M. Rost \orcidlink{0000-0002-8306-1743}}
\affiliation{%
Max Planck Institute for the Physics of Complex Systems, 
Noethnitzer Str. 38, 01187 Dresden, Germany}

\begin{abstract}
We investigate how dipolar interactions between microwave-shielded polar molecules structure the self-bound droplets formed  under variation of the interaction strength. We identify the transition from droplets to crystals as a finite-size first order transition. With  droplet-ring states and transitional supersolid states we predict additional structure in the crystal and droplet phases, respectively.
To describe this strongly correlated regime, and in particular the reconfiguration of quantum ground states, we
 design a variational Monte Carlo framework based on neural quantum states. It is especially suitable to
 describe ground states and  almost degenerate states with very different configurations. Moreover, 
 one can easily determine the superfluid fraction.
Our results reveal the sequence of finite-size structures through which dipolar interactions reorganize molecular droplets into crystals.
\end{abstract}

\maketitle

Polar molecules have reached the quantum-degenerate regime: the first Bose-Einstein condensate (BEC) of strongly dipolar molecules ~\cite{wil24} was realized only recently by substantially suppressing the reactivity between the molecules.
Through microwave dressing of different rotational states, the molecules are prevented from accessing short distances, where chemical reactions occur, thereby yielding long-lived dipolar molecular gases~\cite{doy21,wan23,kar25}.
Compared to atomic dipolar systems~\cite{pfa09,fer23}, microwave-shielded polar molecules (MSPMs) possess much larger dipole moments, allowing dipolar physics to be pushed from the weakly interacting~\cite{pup12,luk14} toward a genuinely strongly correlated regime~\cite{ wil24, wil25}.
This renders MSPMs an ideal platform for exploring collective phenomena in strongly interacting quantum systems~\cite{dem02,vis18} including quantum droplets and supersolids \cite{lan22,rei25,tao25,wan26,zangprx2025,baenaprr2025,zhang2025supersolid,budjemaapra2025,cardinalenatcom2026,ciardi2026equilibrium,melero2026self}.

\begin{figure}[t]
    \centering
    \includegraphics[width=0.95\linewidth]{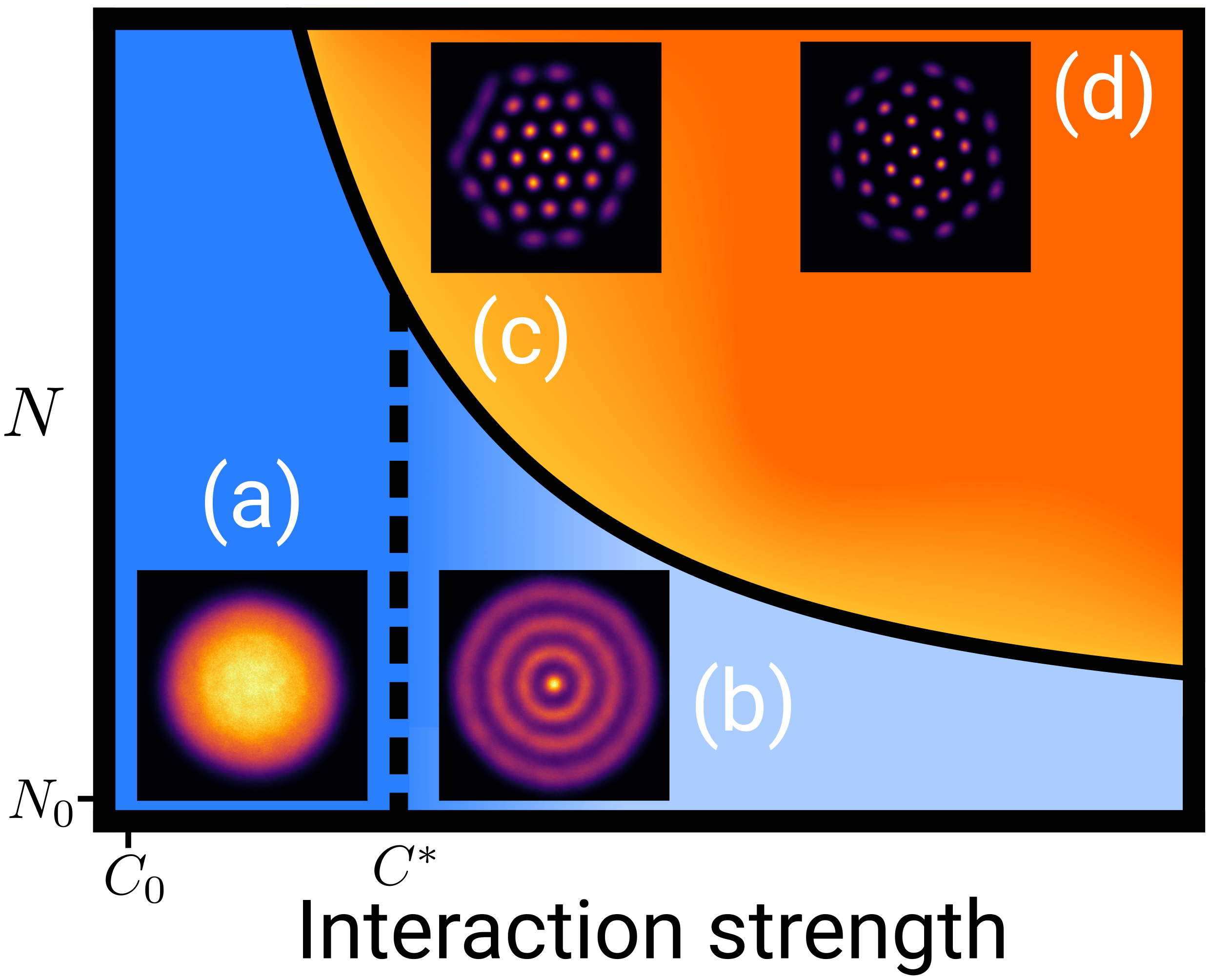}
    \caption{Illustration of ground state geometries with their dependence on system size $N$ and interaction strength $C$. The solid black line marks the transition boundary between the droplet (blue) and crystal (orange) states; color gradients indicate smooth structural changes.  $C^*$ denotes the interaction strength, where the first dimer state occurs. The origins of the axes are given by $N_0=20$ and $C_0=8$. %
    }
    \label{fig:main-res}
\end{figure}

The tunability of the molecular interactions together with the  high molecular densities enabled by the microwave shielding, lead to 
 a rich phase diagram of the MSPMs~\cite{poh26}.
Beyond the well known BEC phase,  self-bound droplets have been established in theory and experiment~\cite{wil24,wil25, langenprl2025,jinprl2025,bailliepra2026}, and a monolayer crystal has  been predicted deep in the strongly interacting regime only recently~\cite{poh25}.
\Cref{fig:main-res} sketches droplet and crystal phases as a function of the number of molecules $N$ and interaction strength $C$, where $C^*$ marks the appearance of the first dimer state. The insets (a-d) provide representative density profiles hinting at more detailed structures beyond the droplet and crystal phases which we will discuss in the following, along with the nature of the droplet-to-crystal transition whose nature is not known to date.

We identify the droplet-to-crystal transition as first-order with abrupt changes in both, superfluid fraction and density-density correlations across the phase boundary. Further, we demonstrate that  droplet and  crystal state energies exhibit a crossing at the phase-boundary, a symmetry-preserving signature of a first-order transition. Analyzing these signatures   in finite-size spectra as well as the build up of  correlations  from the few- to many-body limit  depending on particle number and interaction strength establishes a systematic understanding of the transition~\cite{Sachdev}.

With the variation of particle number $N$ and interaction strength $C$ finite systems may develop intermediate structures that have no direct counterpart in the thermodynamic limit $N{\to}\infty,\, C{=}\text{const}$. They belong to   the microscopic pathway from liquid to crystal.
Indeed, 
near the phase-boundary, two additional size-dependent states emerge: on the droplet side for $C>C^*$, a droplet-ring state with radial density modulations appears [\cref{fig:main-res}(b)], connected to the droplet through a smooth crossover.  On the crystal side, the system displays mixed liquid- and solid-like character [\cref{fig:main-res}(c)], with a nonzero superfluid fraction confined to the edge. The latter resembles a transitional supersolid which recently has been found in ion doped helium systems \cite{gia25}.

In what follows we will  reveal the microscopic pathway from liquid to crystal with a variational Monte Carlo (VMC) approach based on  neural quantum states (NQS)~\cite{fre23,car17}. Overcoming the limited flexibility of  traditional variational wavefunction parameterizations, VMC determined NQS provide near-exact ground state (GS) solutions~\cite{cyb89,fre24,fou24,car25,pfa20,car21,her20,fre25,nag19}, and equally  important in the present context, NQS allow 
us to describe  coexisting liquid- and solid-like states on equal footing.

To be specific, we consider the $N-$body Hamiltonian $\mathcal H=\sum_{i=1}^N K_i + \sum_{i<j}^N V(\vct r_{ij})$ as the sum of kinetic energies $K_i$ of each molecule $i$ and the pairwise intermolecular interactions $V(\vct r_{ij})$ with $\vct r_{ij}=\vct r_i - \vct r_j$.
By dressing molecules' internal rotational states with a single circularly polarized microwave field, the effective intermolecular potential reads \cite{shi25a}

\begin{equation}\label{eq:pot}
V(\vct r) =  E_0 ~C
\left[
\frac{(1 - \cos^4\theta)} {\left(r/R_0\right)^6} +
\frac{(3\cos^2\theta - 1)} {\left(r/R_0\right)^3} 
\right] ~,
\end{equation}
where  $\theta$ is the  angle between the polarization axis and the vector $\vct r$ connecting two dipoles, while $r=|\vct r|$ represents the relative distance. 
Length and energy scales are defined as $R_0=\left({C_6}/{C_3}\right)^{\nicefrac 1 3}$ and $E_0={\hbar^2}/({M R_0^2})$, in terms of the molecular mass $M$, and the potential coefficients $C_3$ and $C_6$. Then,
\begin{equation}
    \label{eq:C}
    C= {C_3^2} / ({C_6 E_0})
\end{equation}
 quantifies the overall strength of the molecule-molecule interaction with   $C_3{=}{d^2\Omega^2}/[{48\pi\epsilon_0(\Omega^2+\Delta^2)}]$ and $C_6{=}{d^4\Omega^2}/[{128\pi^2\epsilon_0^2\hbar(\Omega^2+\Delta^2)^{\nf 3 2}}]$, %
where $d$ is the molecular dipole moment, $\Omega$ is the Rabi frequency, and $\Delta$ is the frequency detuning of the microwave field.

\cref{eq:pot} suggests that the system favors configurations in which the molecules are distributed along the $xy$ plane while their $z$-axis motion is constrained.
We have simulated  system sizes of $N=2-40$ molecules and interaction strengths $C\le 60$ with $C^*=12.3$ within experimental capabilities
\footnote{LiCs molecules~\cite{wei10} ($d\approx5.5$ D and $m\approx140$ amu) can possess interaction strength $C\le60$ for Rabi frequencies $\Omega/ 2 \pi \approx10~\rm{MHz}$ ~\cite{doy21, wan23}.}.
\nocite{wei10}

For $T=0$, the Hamiltonian $\mathcal H$ is addressed in the VMC framework based on an artificial neural network (ANN) \cite{fre23}, where trial states are given by NQS.
More specifically, we propose a NQS %
for MSPMs that provides direct access to the GS while remaining robust against system fragmentation, without the need of auxiliary traps \cite{bor25} or collective cluster moves \cite{poh25}. The trial state is given by
\begin{equation}\label{eq:psi}
\psi(\vct R) = \exp\left[-\sum\limits_{i<j} \frac b {r_{ij}^m} \right]
\sum\limits_\alpha \omega_\alpha \chi_\alpha(\vct R)\,,
\end{equation}
 where the spatial coordinates for all N molecules are encoded by $\vct R=\{\vct r_1,...,\vct r_N\}$, and $m$, $b$ and $\omega_\alpha$ are variational parameters. 
In~\cref{eq:psi}, the sum over $\alpha$ represents a linear combination of many-body states.
The functions 
\begin{equation}\label{eq:orbitals}
\chi_\alpha (\vct R) = \prod_{i=1}^N \sum_{\mu=1}^N h_{\alpha\mu} (\vct r_i,\{\vct r_{i/}\}) \exp[-a_{\alpha\mu} q_i^2]
\end{equation}
are constructed to incorporate the ANN outputs and Gaussian decay factors, 
ensuring the wavefunction is normalizable. It contains additional variational parameters
 $a_{\alpha\mu}$. 
The distance of the $i$-th molecule to the center of mass $\vct r_{\rm {cm}}$ is defined by $q_i=|\vct r_i - \vct r_{\rm {cm}}|$, 
and the coordinates of all molecules excluding $\vct r_i$ are represented by the set $\{\vct r_{i/}\}$.
Through this separation, and by requiring that the ANN outputs $\vct h_i\equiv\left\{h_{\alpha\mu}(\vct r_i,\{\vct r_{i/}\})\right\}$ are made invariant with respect to the ordering of $\{\vct r_{i/}\}$, it is guaranteed that $\chi_\alpha$ is symmetric under exchange of particles by construction. 
For the computation of the ANN outputs $\vct h_i$, the two-stream architecture described in~\cite{fre23} was employed,
where one stream processes the single molecule information and the other one the molecular pair information. Finally, the output of the two streams is composed such that the combined output is invariant with respect to 
the ordering of the $\{\vct r_{i/}\}$.

\begin{figure}[t]
    \centering
    \includegraphics[width=0.99\linewidth]{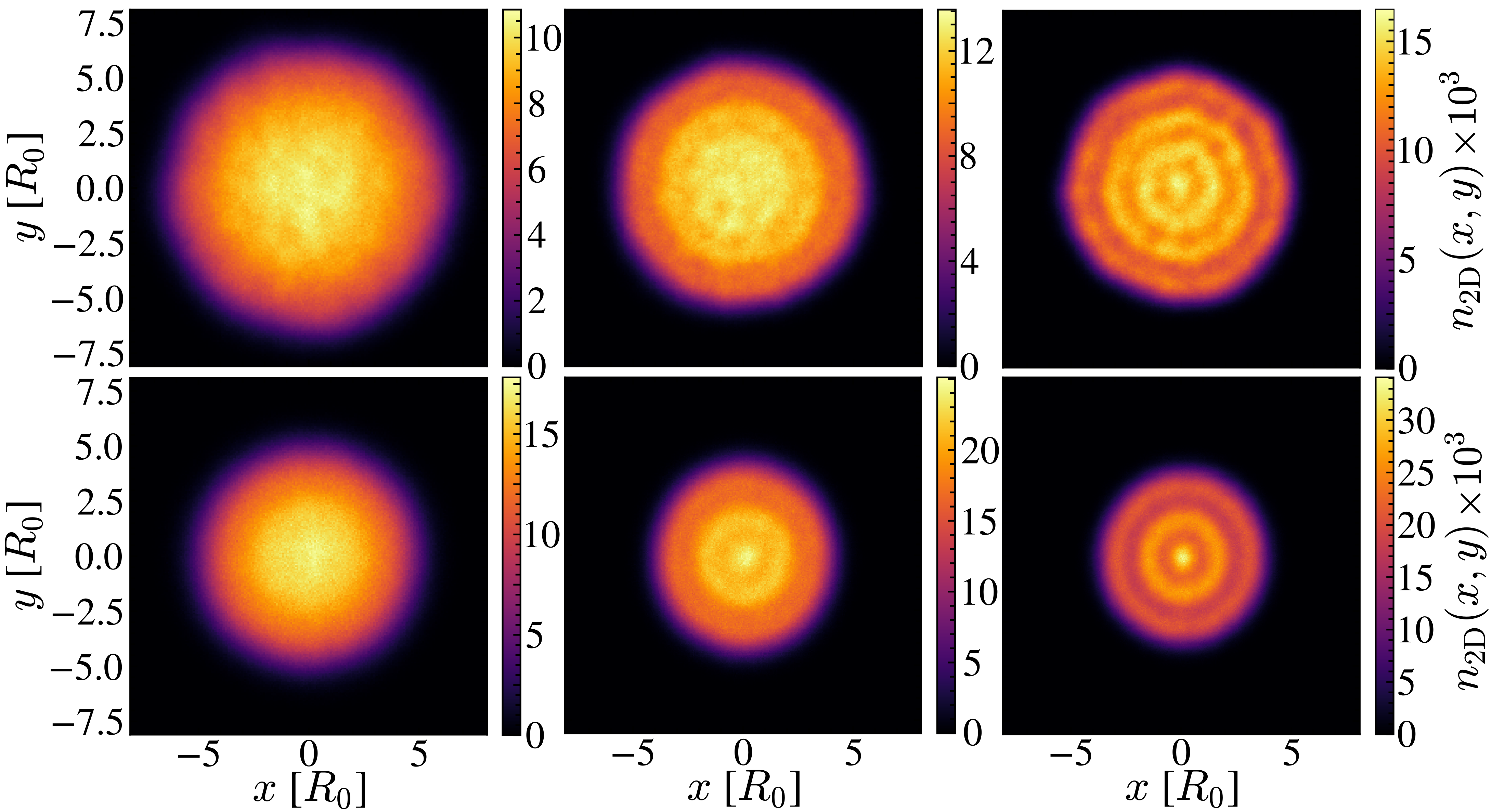}
    \caption{Gradual formation of radial modulations with increasing interaction $C$ for droplets containing    $N=37$ (top row) and $N=19$ (bottom row) molecules. From left to right, the panels correspond to $C=~ 10,~ 15,~\rm{ and}~ 20$. Note that the potential supports at least one dimer state for $C>C^*=12.3$. }
    \label{fig:formation-rings}
\end{figure}

For weak interactions, the system is dilute and the anisotropy in~\cref{eq:pot} mildly affects the geometric shape of the GS which resembles  a nonself-bound  three-dimensional gas-like state, which we do not consider here~\cite{tao25,bor25,poh25}.  

For large $C$, the antidipolar $r^{-3}$ term of \cref{eq:pot} dominates, resulting in the formation of a self-bound many-body state, i.e. a {\it droplet},  see panel (a) in~\cref{fig:main-res}.
In general, the system contracts as $C$ grows at fixed $N$, while it expands with increasing $N$ for fixed $C$, as 
the 2D density profiles in~\cref{fig:formation-rings} reveal.
Moreover, increasing the interaction strength amplifies the anisotropy of the droplet  which ultimately takes on a quasi-two-dimensional shape \cite{poh25}.

For $C$ larger than the critical interaction strength $C^*=12.3$ the crossover is accompanied by the formation of droplet-ring states, see  \cref{fig:main-res}(b). 
\cref{fig:formation-rings} shows in detail the crossover behavior: The 2D density profiles of the droplet gradually develop radial modulations as $C$ increases. 
Hence, the droplet to  droplet-ring state transition is a crossover.
It is physically rooted in the existence of a shallow two-body bound state, as we will discuss now.

From a two-body perspective the corresponding scattering length diverges at $C{=}C^*$, allowing the two-body wavefunction to extend over the droplet size.
In this regime, particle delocalization dominates, yielding a nearly uniform droplet.
However, for increasing $C>C^*$, the scattering length decreases. Once smaller than   the droplet, it defines a new length scale at which pairs of molecules bind at finite separation leading to  regions of amplified density.

\begin{figure}[t]
    \centering
    \includegraphics[width=0.95\linewidth]{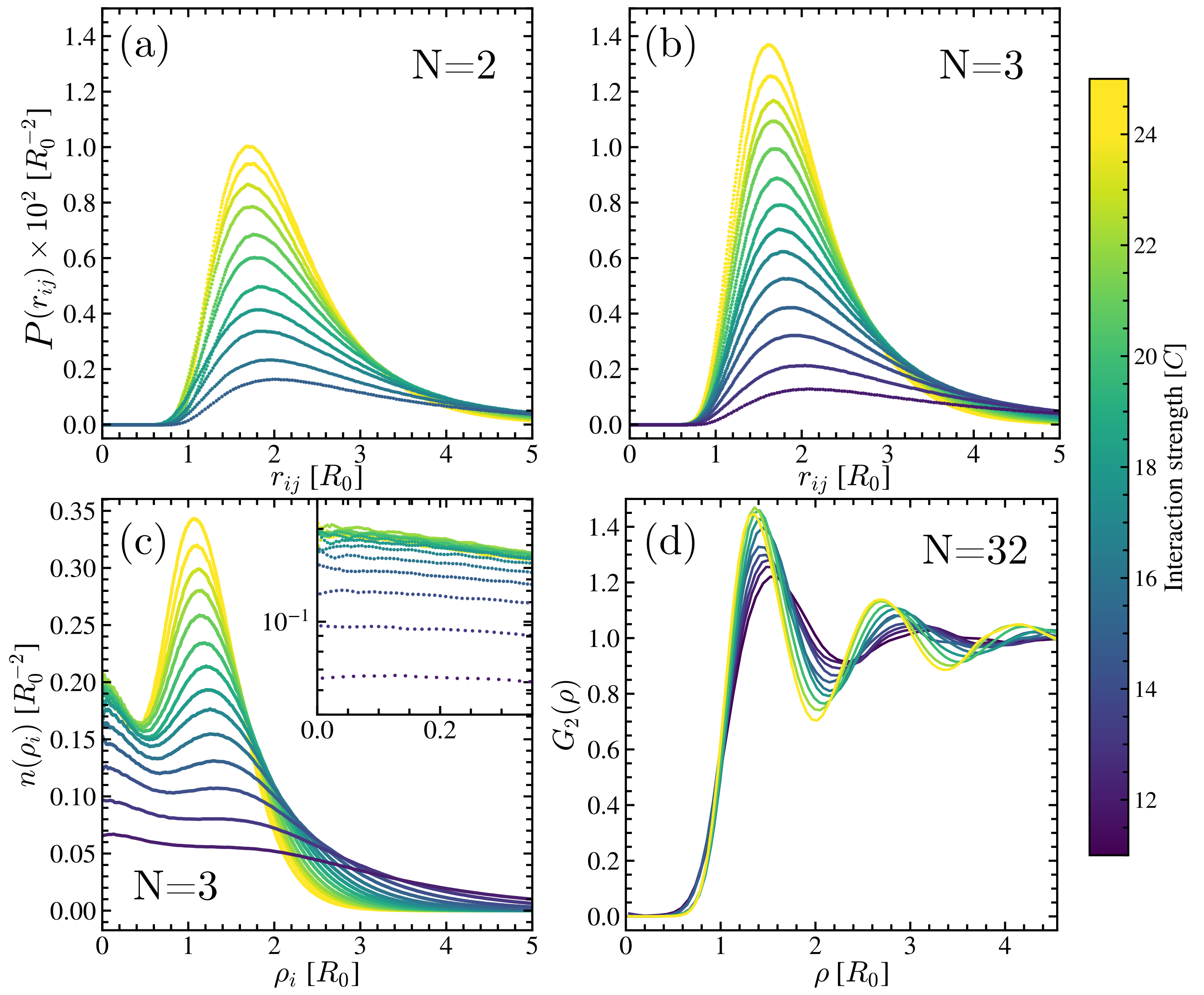}
    \caption{Correlated molecular densities;  (a) and (b):  pair densities $P(r) = \frac 2 {N(N-1)} \left\langle \sum_{i<j} {\delta(r_{ij} - r)}/{r^2}\right\rangle $ as a function of the distance $r_{ij}$ for dimers and trimers, respectively; (c): in-plane density profile $n(\rho) = \left\langle \sum_i {\delta(\rho_i - \rho)}/{\rho}\right\rangle$  as a function of the $xy$ distance to center-of-mass $\rho_i$ for the trimer. The inset in (c) refers to the behavior near the center-of-mass; 
     (d):  density-density correlations $G_2(\rho)$ from \eqref{eq:g2} for $N=32$ molecules as a function of the in plane  relative distance $\rho$. 
    }
    \label{fig:cor-n2-n3}
\end{figure}

On the other hand,  regions of reduced density originate from the repulsive core of the intermolecular potential preventing molecules from approaching each other closely, which can be clearly seen from
the pair density functions  of the dimer $N=2$ shown  in \cref{fig:cor-n2-n3}(a). Note that the 
pair-density peaks around $1.8R_0$ for dimers and trimers (\cref{fig:cor-n2-n3}(b)),  regardless of  the interaction strength $C$. Obviously, the density modulation in the droplet-ring states is dominated by two-body physics.

 The trimer density profiles  develop  radial modulations for $C>C^*$, as can be seen in
  \cref{fig:cor-n2-n3}(c).
The inset of panel (c) demonstrates that the density  of the trimer saturates for large $C$ at the center as a consequence of the  suppression in the pair density for $r<R_0$ due to intermolecular repulsion.
Hence, with one molecule at the origin and the remaining ones located at distances of order $R_0$ leads to density modulations while the overall size of the droplet decreases with increasing interactions strength,
as can be seen from the density tails in  \cref{fig:cor-n2-n3}(c).
Viewed in this way,  droplet-ring states emerge from  a few-body mechanism: the dimer size sets a preferred pair distance which leads in combination with the density saturation to regions of alternating amplified and reduced density. This is, how 
 the droplet to droplet-ring crossover reflects the gradual buildup of few-body correlations.

\begin{figure*}[t]
    \centering
    \includegraphics[width=\textwidth]{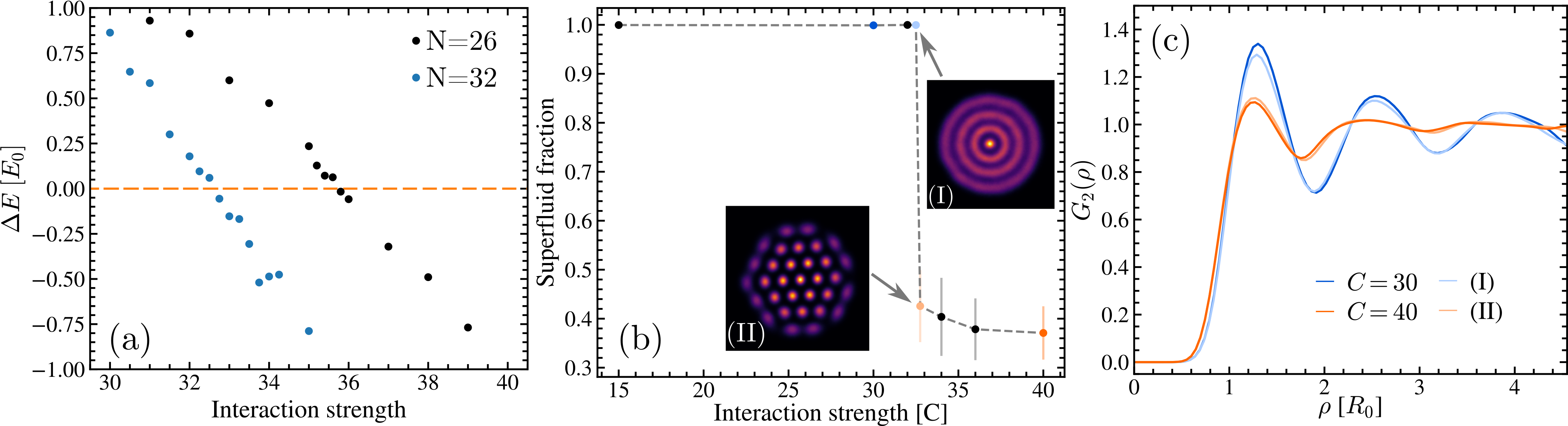}
    \caption{%
    Structural changes across the transition form droplet to crystal as function of interaction strength $C$: (a) difference of droplet-ring and crystal energies from  pair simulation, see text; 
    (b) superfluid fraction (see text) and density profiles immediately before (I) and after (II) the transition for $N{=}32$;  (c) density-density correlation $G_2(\rho)$ from \eqref{eq:g2} for selected $C$ values and $N{=}32$.%
   }
    \label{fig:transition}
\end{figure*}
As a final  signature revealing  the character of the droplet to droplet-ring states transition we analyze
 the density-density correlation function $G_2(\rho)$  for $N=32$  
 in the $xy$ plane.  With one  molecule  at the center of mass position $\rho'{=}0$, it is derived from the general two-body correlation
\begin{align}
    \label{eq:g2}
   G_2(\rho) = \lim_{\rho'\to 0} \frac{\braket{ \sum_{i\neq j} \frac{\delta(\rho_i-\rho)}{\rho} \frac{\delta(\rho_j-\rho')}{\rho'}}}{
   {n(\rho) ~n(\rho')}}\,.
\end{align}
As can be seen in \cref{fig:cor-n2-n3}(d),  the oscillations of $G_2(\rho)$
 change only quantitatively as $C$ increases across $C^*$ confirming the smooth crossover from droplet to droplet-ring states.

Having discussed the crossover to droplet-ring states in the droplet regime,
we come now to  the transition from droplet to crystal states which, in contrast, is discontinuous as \cref{fig:transition} illustrates.
Since both states are self-bound, they are associated with two distinct discrete energy levels of $\mathcal H$, corresponding to the GS and a low-lying excited state.
Hence, as the system evolves from droplet to crystal with increasing $C$, the corresponding energy levels can either cross exactly or form a narrow avoided crossing.
In both cases, the two states have nearly degenerate energies close to the transition boundary.
In order to carefully distinguish them, pairs of simulations were performed with initial conditions favoring droplet or crystal states. 
The initial conditions consist of a pre-optimized trial function together with a set of samples $\vct R$ drawn from the same function. 
For system sizes of $N=32$, the initial conditions of the droplet-ring/crystal state is obtained from the same optimized wavefunction that leads to the density profiles~\cref{fig:main-res} (b)/(c) at interaction strength $C{=}30$/$C{=}40$.
Using these two initial conditions, simulations with $C\in[30,35]$ give the difference $\Delta E = E_c - E_r$ between droplet-ring ($E_r$) and crystal ($E_c$) energy states, where $\Delta E>0$ indicates $E_r$ as the GS. 

From
\cref{fig:transition}(a) one can see that $\Delta E\approx 0$ at $C=32.6$.  Hence, within numerical resolution, the energies of the droplet-ring and cryrstal states are degenerate implying  an exact crossing. Yet,  
droplet-ring and crystal states represent two different geometric configurations, corroborated by their considerably different partition of kinetic and potential energies  despite their same total energy (see {EndMatter, C.}, also verified for different particle numbers $N$). This fact suggests that  the droplet-to-crystal transition is a finite size  first order phase transition.

Further evidence for the nature of the transition is found in structural and correlation observables, which reveal the discontinuity  more clearly.
One of them is the superfluid fraction. We determine it from the linear response of the GS
 to small rotations around the $z$-axis and compute the non-classical rotational inertia (EndMatter, C.). 
In the co-rotating frame, the Hamiltonian of the system is $\mathcal H'=\mathcal H-\omega L_z$, for which the corresponding GS is obtained and used to compute the system's superfluid fraction through $f_s=1-\lim\limits_{\omega\to0} \langle L_z \rangle/(\omega I_{\rm cl})$ with   the classical momentum of inertia $I_{\rm cl}$.
Traditional trial functions within VMC calculations are insufficient to provide reliable results for the superfluid fraction.
 Exploiting the flexibility of NQS, however, near-exact solutions of $\mathcal H'$ can be obtained
  (see {EndMatter, B.}).

\cref{fig:transition}(b) shows that the superfluid fraction $f_s$ 
changes discontinuously at the droplet-to-crystal transition  $C{=}32.6$. This is also the case for  the density profiles at the center of mass  (see {EndMatter, C.}).
Similarly, the density-density correlation $G_2(\rho)$  undergoes an abrupt change, as can be seen in \cref{fig:transition} (c):  the contrast of the  $G_2(\rho)$ oscillations due to droplet-ring states (see blue lines) is significantly and suddenly reduced entering the crystal phase  (see orange lines).

Together, the near degeneracy of the droplet-ring and crystal energies, along with the abrupt changes in the superfluid fraction and $G_2$ across the transition boundary, can be interpreted as finite-size signatures of a first-order liquid-to-solid phase transition~\cite{Sachdev, vol23}.

\cref{fig:transition}(b) reveals another subtle feature, namely that even after transition to the crystal states a small superfluid fraction survives.
It originates from the molecules in the outer ring which remain delocalized, as can be seen in panel (II), while the molecules in the inner rings  are well localized.
Therefore, near the boundary, these many-body bound states resemble  transitional supersolids predicted in ion-doped helium droplets, where crystalline and liquid layers are connected by an intermediate supersolid one~\cite{gia25}. 
Here, this type of supersolid state emerges from the just crystallized droplet at the interface to the vacuum. 
In conclusion, we have established a variational Monte Carlo framework based on neural-network quantum states for strongly correlated finite quantum matter, strictly in the ground state. It allows us to uncover and understand structural changes in  finite strongly correlated systems varying particle number and interaction strength as demonstrated here with droplets formed from microwave shielded polar molecules.

Our neural quantum states remove the restrictions of conventional trial wavefunctions enabling  an accurate ground state, while being stable against system fragmentation.
Building on this capability, we have developed a linear-response scheme to small rotations, allowing us to extract the superfluid fraction and use it as a sensitive order parameter to distinguish emergent quantum states.
More specifically, our analysis has shown that the droplet-to-crystal transition exhibits finite size characteristics of a first-order phase transition. 
On the crystal side of the transition  near the droplet-to-crystal boundary, we observe states with small but finite superfluid fraction signaling transitional supersolidity.
On the droplet side of the transition, we predict droplet-ring states whose origin traces back to two-body physics. The transition from droplet to droplet-ring states has the character of  a smooth crossover.

Quantitatively, the  phenomena observed are mostly determined by the form of the interaction and the self-bound character of the system. 
Qualitatively, structures such as droplet-rings and transitional supersolid states are expected to occur for molecular systems with strongly anisotropic interactions~\cite{wil25,wan26}. 

\begin{acknowledgments}
    We thank Dr.~Juan Carlos Acosta Matos and Dr.~Marin Bukov for useful discussions.
\end{acknowledgments}
\bibliography{ref}

\makeatletter 
\renewcommand{\thefigure}{S\@arabic\c@figure}
\makeatother

\makeatletter 
\renewcommand{\theequation}{S\@arabic\c@equation}
\makeatother

\setcounter{equation}{0}
\setcounter{figure}{0}

\section*{End matter}

\subsection{A. Additional trial function details}

\begin{figure}[b!]
    \centering
    \includegraphics[width=0.99\linewidth]{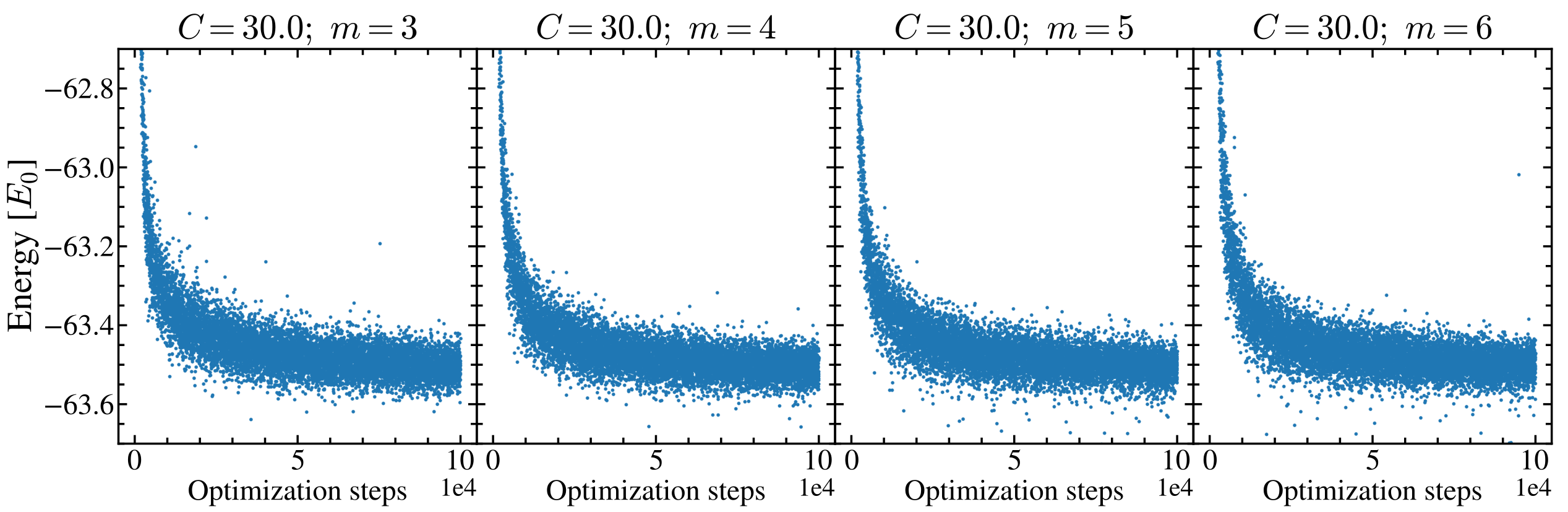}
    \caption{Optimization process showing the evolution of the energy as a function of the iterations for a system with $N=12$ molecules and $C=30$. Panels from left to right correspond to results for $m=$ 3, 4, 5, and 6, respectively.}
    \label{fig:m-opt}
\end{figure}

Employing an isotropic hard sphere  to stabilize the optimization~\cite{fre23}, is not viable due to the anisotropy of the interaction potential in~\cref{eq:pot}. The simulations presented here do not use any kind of hard sphere to stabilize the optimization process. Consequently, %
the asymptotic $r\to 0$ solution for the $1/{r^k}$ potential 
requires that $k=2m+2$ in~\cref{eq:psi}.
Simulations  
with $m=2$, i.e. the standard exponent for power law potentials with $k=6$ as \cref{eq:pot}, featured several instabilities making the optimization procedure unfeasible. 
In fact, the anisotropic character of the potential confers to the interaction an effective extra repulsion. Acknowledging to this fact,  simulations were performed with integer $m \in [3,6]$. Results for optimization energies as a function of the iterations for $N=12$ molecules and interaction strength $C=30$ are displayed in~\cref{fig:m-opt}.

Within this range of $m$ results were insensitive to the actual value of $m$, although $m=5$ proofed  to be the best alternative and is routinely used  in this work. 
Taking this solution, the simulations are stable for the range of interaction strengths examined. 

Regarding the size of ANN, the typical hyperparameters employed consisted on a batch size of 4096 configurations, 4 layers with single molecule stream width of 256, a molecular pair stream width of 32, and 16 functions $\chi_\alpha$. After the energy minimization through the variational Monte Carlo method, estimation of the desired observables are done by fixing the variational parameters and performing Monte Carlo integration of the relevant quantities. %

\subsection{B. Computing the superfluid fraction} 

Suppose that a small rotation with angular frequency $\omega$ according to the $z$ axis is applied to the system. The rotating system is described by $\mathcal H' = \mathcal H - \omega L_z$ when writing the Hamiltonian in the co-rotating frame~\cite{fet08}. Denote the GS of the non perturbed system as $\psi_0(\vct R)$. Since the perturbation is purely rotational, the dominant effect at small $\omega$ is the emergence of a phase twist in the many-body wavefunction~\cite{leg98}, while the density probability $\psi_0^2(\vct R)$ remains approximately unchanged. Motivated by this observation, we make the variational ansatz
\begin{equation}
\Psi (\vct R) = \psi_0(\vct R) ~e^{i\phi(\vct R)} ~,
\end{equation}
where 
\begin{equation}
\phi(\vct R) = \frac \pi M \sum_{\mu=1}^M \sin\left[h'_\mu\left(\vct r_i, \{\vct r_{i/}\}\right)\right]
\end{equation}
is the phase function and $h'_\mu$ are ANN outputs using the same two-stream architecture as $h_{\alpha\mu}$. Obviously, a different set of variational parameters is used for $h'_\mu$.

Since the trial function in~\cref{eq:psi} is based on neural networks, the optimized function for the non-perturbed system is a good approximation for $\psi_0$. Hence, by employing an already optimized trial wavefunction as $\psi_0$, only the variational parameters of $\phi$ need to be optimized to find the optimal phase. To this end, the  functional 
\begin{equation}
\Delta E = \int d\vct R ~p(\vct R) \sum_{i=1}^N \frac{\left[\nabla_i \phi(\vct R)\right]^2} 2 - \hbar\omega \left[\vct r_i \times \nabla_i\phi(\vct R)\right]_z
\end{equation}
has to be minimized, 
 where $p(\vct R) \propto |\psi_0(\vct R)|^2$ is the density probability generated by the unperturbed GS.
For updating the variational parameters of the phase $\phi$, the ADAM~\cite{adam} algorithm was employed. 

After performing simulations for rotation frequencies per particle of $\hat\omega=0.02, 0.04, 0.06, 0.08,$ and $1$, corresponding to $\hbar\omega=\hat\omega E_0/ N$, the angular momentum linear response can be computed.
For each simulation, the expectation values of angular momentum $\langle L_z\rangle$ and the classical inertia tensor $I_{\rm cl}=\langle \sum_i x_i^2+y_i^2\rangle$ are estimated. The superfluid fraction can be calculated by averaging $1-\langle L_z \rangle/(\omega I_{\rm cl})$ over rotational frequencies.

\begin{figure}[t]
    \centering
    \includegraphics[width=0.95\linewidth]{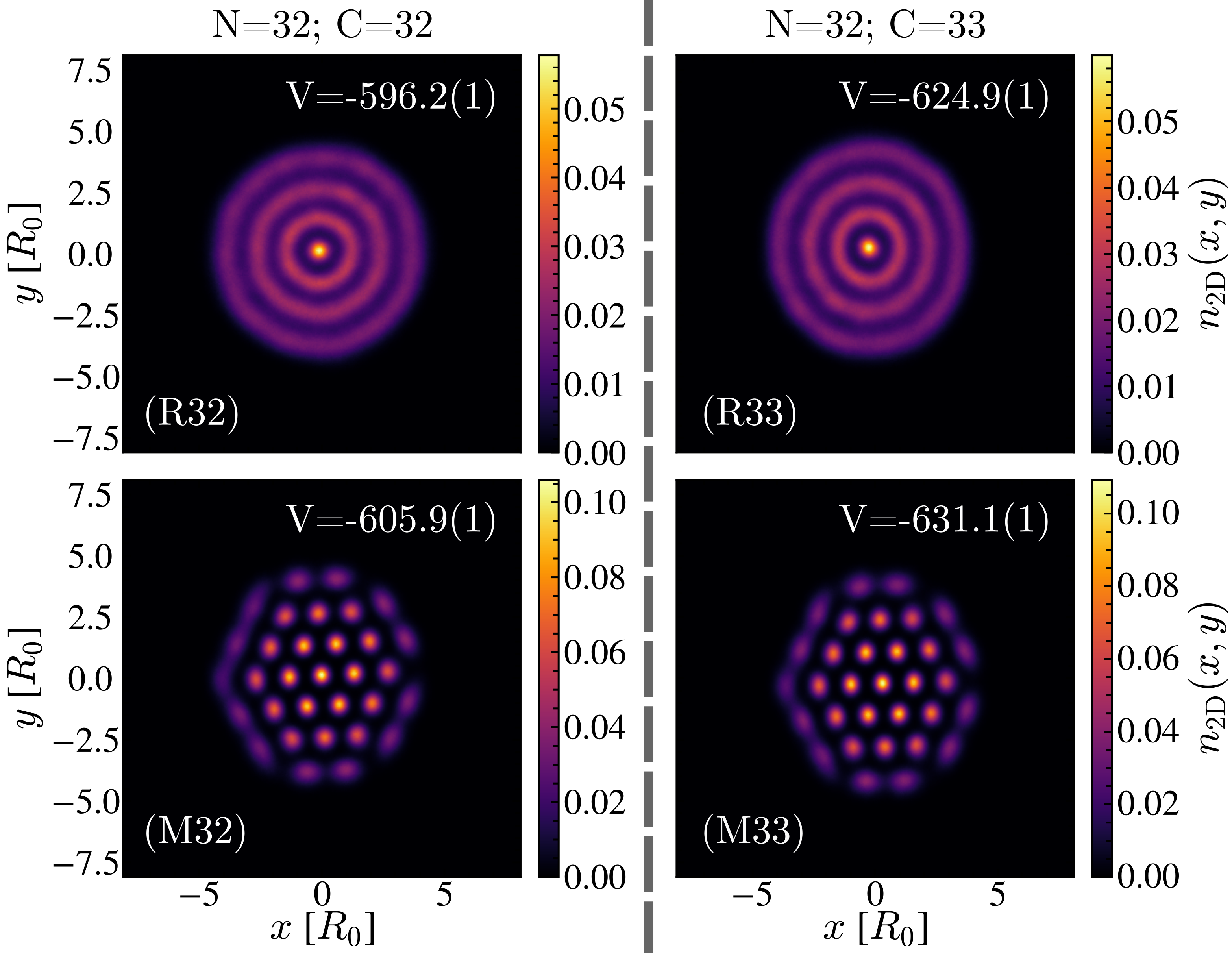}
    \caption{Density profiles for pairs of simulations  with different initial conditions before/after ($C{=}32/C{=}33)$ the transition.
    Panels (R32) and (R33) correspond to initial conditions favoring ring states, while panels (M32) and (M33) favor crystal states.}
    \label{fig:states-diff}
\end{figure}

\begin{figure}[ht]
    \centering
    \includegraphics[width=0.85\linewidth]{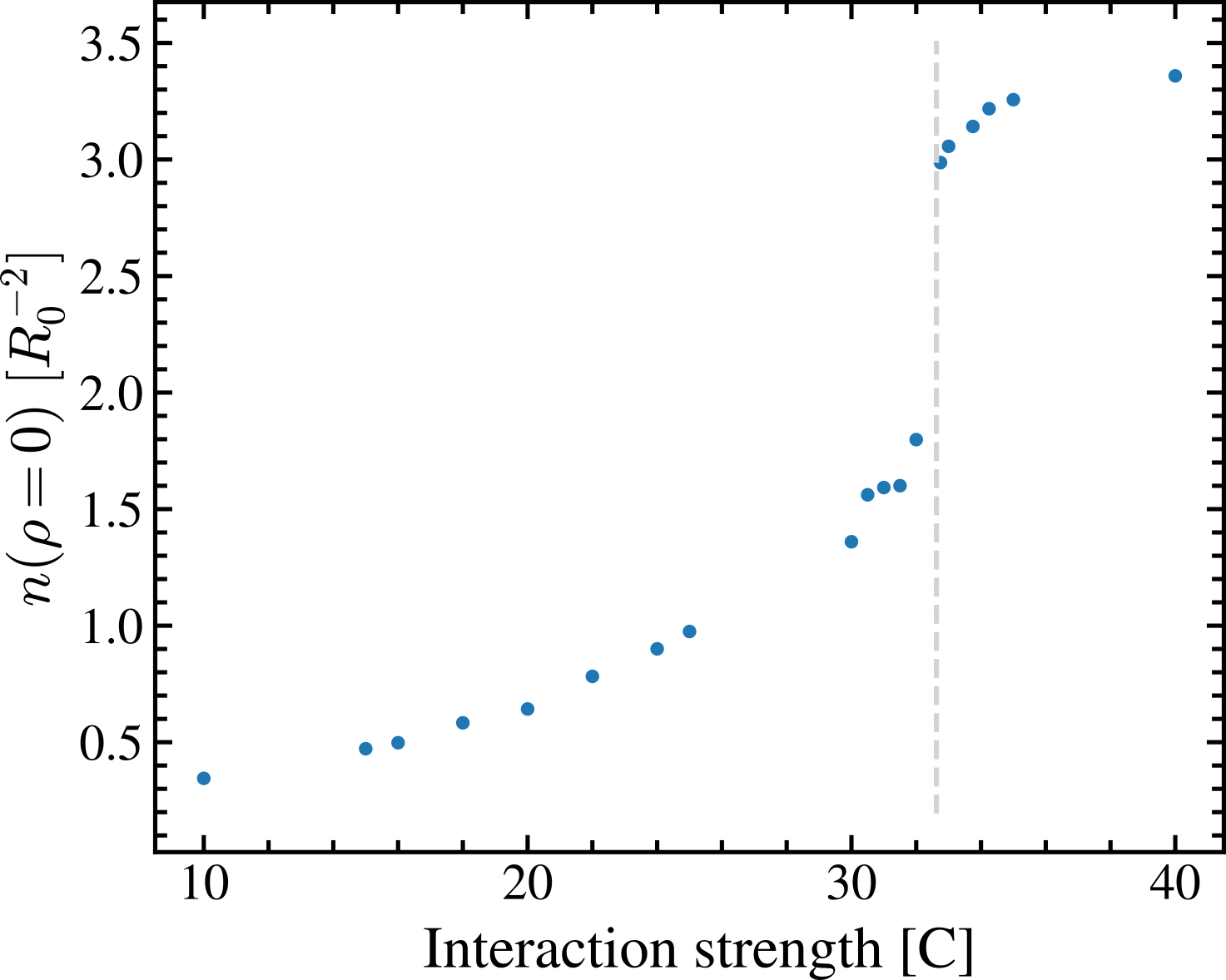}
    \caption{Density profile at the center of mass for the $\rho=\sqrt{x^2+y^2}$ coordinate as a function of the interaction strength. }
    \label{fig:cm-density}
\end{figure}

\subsection{C. Properties near the energy level crossing}

As discussed in the main text, the difference $\Delta E = E_c - E_r$ between droplet-ring ($E_r$) and crystal ($E_c$) energy states is virtually zero  near the transition.
To demonstrate that the optimized trial functions represent distinct states,
\cref{fig:states-diff} presents 2D density profiles before and after the transition. 
The initial conditions for the panels (R32) and (R33) was a pre-optimized ring state, as described in the main text. 
Panels (M32) and (M33) had initial conditions based on a crystal state. 
The expectation values of the potential energy are also shown for each panel. 
Comparing values for equal $C$, the quantitative differences demonstrate that the optimization starting from different initial conditions resulted in distinct states.

Once the GS is defined by the pair simulation procedure, computing properties beyond the energy can help identifying the character of the transition.
Particularly, \cref{fig:cm-density} presents the in-plane density profile $n(\rho)$ at the center of mass $\rho=\rho_{\rm CM}$ as a function of $C$. 
Since the system has global translation symmetry, all properties are computed relative  to the center of mass, which is equivalent to consider $\rho=0$ in this case. 
Within the current statistical accuracy of the simulations, $n(\rho)$ is discontinuous  at the transition, similarly as the superfluid fraction discussed in the main text. 
This hints once more at an exact crossing between the energy levels of the droplet-rings and crystal states.

\begin{figure}[ht]
    \centering
    \includegraphics[width=0.95\linewidth]{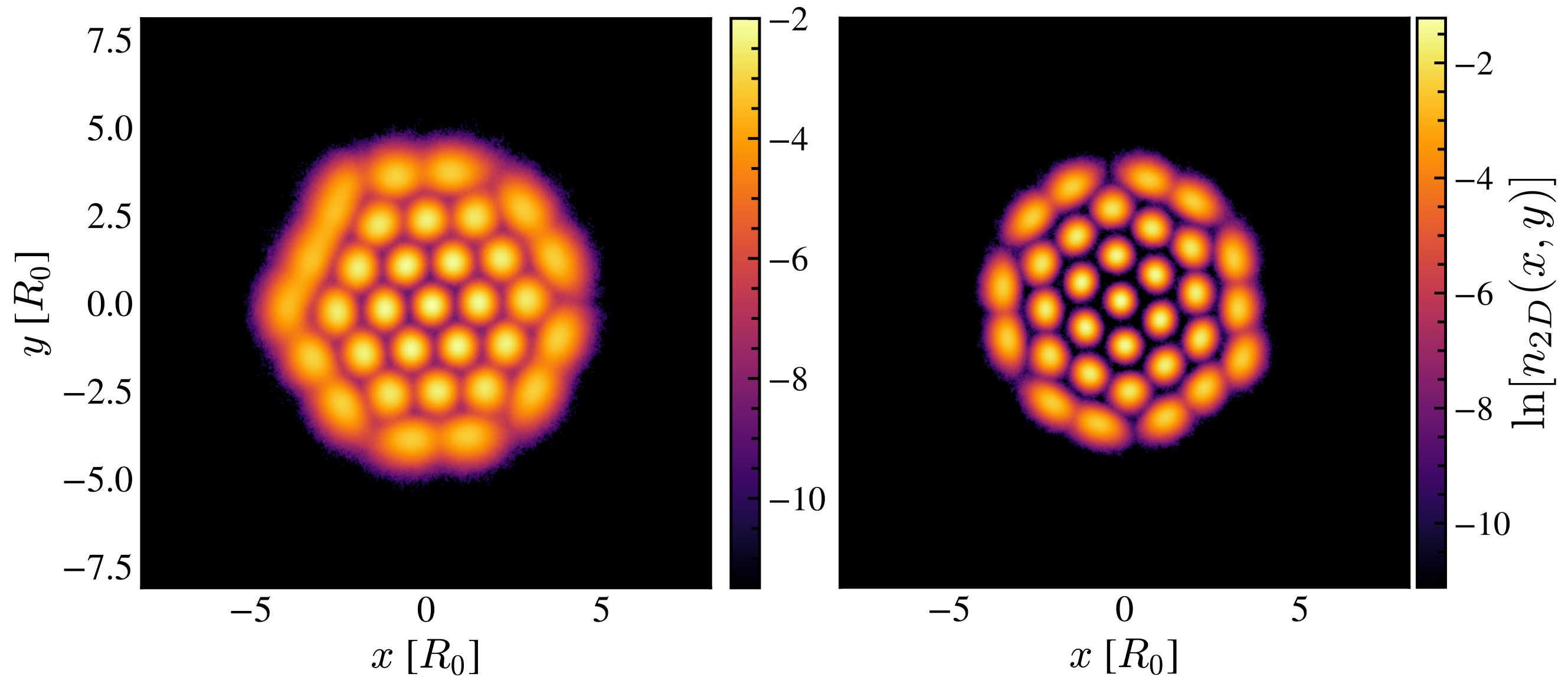}
    \caption{Projected 2D density profiles for a crystal state composed by $N=32$. The left panel shows results for $C=40$ and the right for $C=60$. }
    \label{fig:log-density}
\end{figure}

Right after the transition, crystal states still contain a finite superfluid fraction as pointed out in the main text. The superfluid fraction is hosted by the outermost ring, with some degree of delocalization in the 2D density profile.
This can be clearly seen in
\cref{fig:log-density}. For $C=40$ (left)  the density of the outer molecules overlaps, while there is no overlap for stronger interaction.

\end{document}